\documentclass[conference]{IEEEtran}

\usepackage{cite}
\usepackage{amsmath,amssymb}
\usepackage{bm}
\usepackage{enumitem}
\usepackage{graphicx}
\usepackage[hidelinks]{hyperref}
\usepackage{pgfplots}
\pgfplotsset{compat=1.18}

\newtheorem{lemma}{Lemma}
\newtheorem{proposition}{Proposition}

\newcommand{\btheta}{\bm{\theta}}
\newcommand{\bphi}{\bm{\phi}}
\newcommand{\bdelta}{\bm{\delta}}
\newcommand{\bgamma}{\bm{\gamma}}
\newcommand{\ba}{\mathbf a}
\newcommand{\bv}{\mathbf v}
\newcommand{\bg}{\mathbf g}
\newcommand{\bH}{\mathbf H}
\newcommand{\bC}{\mathbf C}
\newcommand{\bJ}{\mathbf J}
\newcommand{\bR}{\mathbf R}
\newcommand{\bx}{\mathbf x}
\newcommand{\by}{\mathbf y}
\newcommand{\bz}{\mathbf z}
\newcommand{\bn}{\mathbf n}
\newcommand{\bs}{\mathbf s}
\newcommand{\be}{\mathbf e}
\newcommand{\bI}{\mathbf I}
\newcommand{\bSigma}{\boldsymbol{\Sigma}}
\newcommand{\bbR}{\mathbb R}
\newcommand{\E}{\mathbb E}
\newcommand{\cS}{\mathcal S}
\newcommand{\CN}{\mathcal{CN}}
\newcommand{\tr}[1]{\operatorname{tr}\!\left(#1\right)}
\newcommand{\diag}{\operatorname{diag}}

\begin{document}

\title{A Two-Term Ziv--Zakai Bound for Multitarget DoA Estimation in MIMO Radar}

\author{\IEEEauthorblockN{Mohammadreza Bakhshizadeh Mohajer\IEEEauthorrefmark{1}, Daniela Tuninetti\IEEEauthorrefmark{2}, and Luca Barletta\IEEEauthorrefmark{1}}
\IEEEauthorblockA{\IEEEauthorrefmark{1}Department of Electronics, Information and Bioengineering, Politecnico di Milano, Milan, Italy\\
\IEEEauthorrefmark{2}Department of Electrical and Computer Engineering, University of Illinois Chicago, Chicago, IL, USA\\
Email: \{mohammadreza.bakhshizadeh, luca.barletta\}@polimi.it, danielat@uic.edu}}

\maketitle

\begin{abstract}
Multitarget direction of arrival (DoA) estimation is challenging
when fluctuating targets are closely spaced: local
Cram\'er--Rao bounds (CRBs) miss global errors, while direct
vector Ziv--Zakai bound (ZZB) evaluation is computationally
demanding. For colocated multiple-input multiple-output (MIMO)
radar with Swerling-II fluctuations and an ordered angular prior,
we derive a rigorous two-term Bayesian mean squared error lower
bound. A finite separation contribution addresses nonlocal
discrimination, while a correction based on the prior averaged
Fisher information matrix (FIM) retains target coupling and
virtual array geometry. We show that the prior averaged
pointwise CRB diverges under unrestricted target spacing when
the FIM is nonsingular almost everywhere, although the proposed
bound remains finite. Once the averaged FIM is obtained,
evaluation avoids displacement optimization and multidimensional
binary error integration. The bound supports analysis of
snapshot diversity, target spacing, and transmit and receive
allocation. Numerical comparison with the vector ZZB shows
close agreement at low SNR but greater conservatism at high SNR,
quantifying the tradeoff in tightness.
\end{abstract}

\section{Introduction}
\label{sec:introduction}

Colocated multiple-input multiple-output (MIMO) radar exploits transmit waveform diversity and receive spatial processing for direction of arrival (DoA) estimation~\cite{Hassanien2011,StoicaLiXie2007}. With multiple fluctuating targets, covariance coupling and closely spaced directions complicate performance assessment. The Cram\'er--Rao bound (CRB) characterizes local accuracy but misses large errors and threshold behavior~\cite{Athley2005,AbramovichJohnson2010,AmarWeiss2008}. The Ziv--Zakai bound (ZZB) instead bounds Bayesian mean squared error (MSE) through binary tests between distinct parameter values~\cite{ziv1969some}.

Vector and bearing ZZBs were developed by Bell \emph{et al.}~\cite{EZZB_Bell_vect,EZZB_Bell_DoA}; multitarget DoA bounds subsequently addressed permutation ambiguity~\cite{EZZB_1D_DoA,EZZB_2D_DoA}. Related analyses include explicit ZZBs for sparse linear arrays~\cite{KhanBell2013}, a single-target Weiss--Weinstein bound for colocated MIMO radar~\cite{Tran2012_WWB_MIMO}, and ZZBs for joint parameter estimation and distributed MIMO radar~\cite{Chiriac2015_ZZB_MIMO,Liang2024_ZZB_Distributed_MIMO}. For colocated MIMO radar with Swerling-II fluctuations and an ordered angular prior, direct multitarget vector ZZB evaluation entails multidimensional prior integration, displacement optimization, and repeated binary tests. Moreover, when arbitrarily close directions are permitted, averaging the pointwise CRB can be nonintegrable. This motivates a finite, rigorous Bayesian bound that retains nonlocal error information and target coupling without direct vector ZZB evaluation.

We derive a two-term relaxation of the coordinate vector ZZB. Its finite separation term provides a nonlocal error guarantee, while a correction using the inverse prior averaged Fisher information matrix (FIM) retains coupling and virtual array geometry. We prove that separation information is $O(D_i^2)$ as adjacent spacing $D_i$ vanishes, implying divergence of the averaged pointwise CRB under the stated nonsingularity condition, although the proposed bound remains finite. Once the averaged FIM is available, the bound avoids displacement optimization and multidimensional binary error integration. Its explicit structure enables comparisons of snapshot diversity, spacing assumptions, and transmit and receive allocation. A numerical vector ZZB comparison shows close low SNR agreement and a high SNR gap, making the tightness tradeoff explicit.

Bold symbols denote vectors and matrices; $\bI_n$, $\be_k$, $(x)_+$, $\mathbf1_{\mathcal A}$, $\otimes$, and $\CN(\boldsymbol\mu,\mathbf R)$ denote the identity matrix, $k$th canonical vector, $\max\{x,0\}$, indicator of the set $\mathcal A$, Kronecker product, and proper complex Gaussian distribution, respectively.

\section{System Model}
\label{sec:system_model}

Consider a colocated narrowband MIMO radar with $N$ transmit
and $M$ receive antennas arranged in aligned uniform linear
arrays (ULAs) with half-wavelength spacing.
The $K$ far-field target directions are
$\btheta=[\theta_1,\ldots,\theta_K]^T$.
For $\theta\in[-\pi/2,\pi/2]$, the receive and transmit
steering vectors are
\begin{align}
\ba(\theta)
&=[1,e^{j\pi\sin\theta},\ldots,e^{j\pi(M-1)\sin\theta}]^T,
\nonumber\\
\bv(\theta)
&=[1,e^{j\pi\sin\theta},\ldots,e^{j\pi(N-1)\sin\theta}]^T.
\label{eq:ula_manifolds}
\end{align}
Let $\bs(t)=[s_1(t),\ldots,s_N(t)]^T$ collect orthonormal
transmit waveforms over $\mathcal T$, satisfying
$\int_{\mathcal T}\bs(t)\bs^H(t)\,dt=\bI_N$.
At snapshot $l\in\{1,\ldots,L\}$, the received signal is
\begin{align}
\by_l(t)
=
\sum_{k=1}^K
\ba(\theta_k)\gamma_{k,l}\bv^T(\theta_k)\bs(t)
+\bz_l(t),
\label{eq:received_signal}
\end{align}
where $\bz_l(t)$ is spatially and temporally white proper
complex Gaussian receiver noise, independent across snapshots.
Under the Swerling-II model~\cite{Swerling1960}, the
scattering coefficients are constant within each snapshot
and satisfy
$\gamma_{k,l}\overset{\mathrm{i.i.d.}}{\sim}
\CN(0,\sigma_\gamma^2)$
across targets and snapshots, independently of the noise.

Define the virtual array steering vector
$\bg(\theta)\in\mathbb C^{MN}$ and its steering matrix
$\bH(\btheta)\in\mathbb C^{MN\times K}$ by
\begin{align}
\bg(\theta)
&\triangleq\bv(\theta)\otimes\ba(\theta),&
\bH(\btheta)
&\triangleq[\bg(\theta_1),\ldots,\bg(\theta_K)].
\label{eq:mimo_signature}
\end{align}
Columnwise vectorization of the matched filter outputs gives
$\bx_l\triangleq
\operatorname{vec}(\int_{\mathcal T}
\by_l(t)\bs^H(t)\,dt)
=\bH(\btheta)\bgamma_l+\bn_l$,
where $\bgamma_l=[\gamma_{1,l},\ldots,\gamma_{K,l}]^T$
and
$\bn_l\overset{\mathrm{i.i.d.}}{\sim}
\CN(\mathbf0,\sigma_n^2\bI_{MN})$.
With known variances $\sigma_\gamma^2$ and $\sigma_n^2$,
let $\eta\triangleq\sigma_\gamma^2/\sigma_n^2$.
Marginalizing $\bgamma_l$ gives
\begin{align}
\bC(\btheta)
&\triangleq\bI_{MN}+\eta\bH(\btheta)\bH^H(\btheta),&
\bx_l|\btheta
&\overset{\mathrm{i.i.d.}}{\sim}
\CN(\mathbf0,\sigma_n^2\bC(\btheta)).
\label{eq:marginal_covariance}
\end{align}

Since the likelihood in~\eqref{eq:marginal_covariance}
is invariant under target permutations, we restrict
$\btheta$ to
$\cS=\{\btheta:\theta_{\min}\le\theta_1\le\cdots
\le\theta_K\le\theta_{\max}\}$,
where
$[\theta_{\min},\theta_{\max}]\subseteq[-\pi/2,\pi/2]$
and $\zeta\triangleq\theta_{\max}-\theta_{\min}>0$.
The ordered uniform prior is
\begin{align}
f_{\btheta}(\btheta)
=
\frac{K!}{\zeta^K}\mathbf1_{\cS}(\btheta).
\label{eq:ordered_prior}
\end{align}
No minimum target spacing is imposed.
We study
$\mathrm{MSE}\triangleq
K^{-1}\E\|\widehat{\btheta}-\btheta\|_2^2$,
averaged over the prior and observations.

\section{Two-Term Ziv--Zakai Bound}
\label{sec:zzb}

\begin{proposition}[Two-term ZZB lower bound]
\label{prop:two_term}
Under the model in Section~\ref{sec:system_model}, assume
$\overline{\bJ}_L\succ\mathbf0$. Then
\begin{align}
    \mathrm{MSE}
    &\ge B_{\mathrm{ZZB}}
    \triangleq\frac{\zeta^2P_\perp}{(K+1)(K+2)}
    +\frac1K\sum_{k=1}^KI_k^{\rm G}.
    \label{eq:final_bound}
\end{align}
\end{proposition}
Here $P_\perp$ is the uniform binary error lower bound,
$\overline{\bJ}_L$ is the prior averaged FIM, and
$I_k^{\rm G}$ is the local correction, defined in
\eqref{eq:uniform_binary}, \eqref{eq:average_fim}, and
\eqref{eq:local_correction}, respectively.
The finite separation and local contributions are derived
from the coordinate vector ZZB below; proofs are provided
in the Appendix.

\subsection{Coordinate Form of the Vector ZZB}
\label{subsec:coordinate_zzb}

Let
\begin{align}
    \bR_e
    \triangleq
    \E\!\left[
    (\widehat{\btheta}-\btheta)
    (\widehat{\btheta}-\btheta)^T
    \right]
\end{align}
denote the error second-moment matrix, so that
$\mathrm{MSE}=K^{-1}\sum_{k=1}^K\be_k^T\bR_e\be_k$.
For a displacement $\bdelta\in\bbR^K$, define the overlap
of the prior supports as
$\cS_{\bdelta}\triangleq
\{\bphi\in\cS:\bphi+\bdelta\in\cS\}$.
Let $P_e(\bphi,\bphi+\bdelta)$ denote the minimum error
probability for testing $\btheta=\bphi$ against
$\btheta=\bphi+\bdelta$ with equal hypothesis probabilities,
based on the $L$ snapshots in~\eqref{eq:marginal_covariance}.
For the ordered uniform prior, applying the vector ZZB
of~\cite{EZZB_Bell_vect} along the coordinate direction
$\be_k$, while omitting the valley filling operation,
yields
\begin{align}
    \be_k^T\bR_e\be_k
    \ge
    \int_0^\infty h
    \max_{\bdelta:\,\be_k^T\bdelta=h}
    \int_{\cS_{\bdelta}}
    f_{\btheta}(\bphi)
    P_e(\bphi,\bphi+\bdelta)
    \,d\bphi\,dh.
    \label{eq:coordinate_zzb}
\end{align}
The valley filling operation can only increase the bound, so its omission
preserves the inequality. The coordinate projections allow
these inequalities to be averaged directly to bound the MSE.
The maximization in~\eqref{eq:coordinate_zzb} is over all
$\bdelta$ satisfying $\delta_k=h$, not only
$\bdelta=h\be_k$; this freedom permits different feasible
displacements for the finite separation and local contributions.

\subsection{Finite Separation Binary Bound}
\label{subsec:nonlocal}

For $0\le h\le\zeta$, choose $\bdelta=h\be_k$, which changes
only target $k$, and write $\vartheta\triangleq\phi_k$ and
$\varphi\triangleq\phi_k+h$. The unchanged targets contribute
an independent Gaussian component with the same distribution
under both hypotheses. Adding this component to the isolated
target observation cannot improve binary discrimination.
Let $P_e^{\mathrm{iso}}(\vartheta,\varphi)$ denote the minimum
error probability for the isolated target test. After
normalization by $\sigma_n^2$, its covariance matrices are
\begin{align}
    \bSigma_i
    &=\bI_{MN}+\eta\bg_i\bg_i^H,
    &\bg_0&\triangleq\bg(\vartheta),\quad
    \bg_1\triangleq\bg(\varphi),
    \label{eq:isolated_pair}
\end{align}
for $i\in\{0,1\}$. These matrices differ from the identity
only on the subspace spanned by $\bg_0$ and $\bg_1$.
Their spatial correlation is
\begin{align}
    \psi(\vartheta,\varphi)
    &\triangleq\bg^H(\vartheta)\bg(\varphi)
    \nonumber\\
    &=\ba^H(\vartheta)\ba(\varphi)\,
    \bv^H(\vartheta)\bv(\varphi).
    \label{eq:spatial_kernel}
\end{align}
For $\Delta\triangleq\sin\varphi-\sin\vartheta$ and
$P\in\{M,N\}$, define
\begin{align}
    D_P(\Delta)
    &\triangleq\sum_{p=0}^{P-1}e^{j\pi p\Delta}
    \nonumber\\
    &=e^{j\pi(P-1)\Delta/2}
    \frac{\sin(P\pi\Delta/2)}
         {\sin(\pi\Delta/2)},
    \quad D_P(2q)=P,\ q\in\mathbb Z.
    \label{eq:dirichlet_kernel}
\end{align}
Consequently,
\begin{align}
    |\psi(\vartheta,\varphi)|
    &=|D_M(\Delta)D_N(\Delta)|,
    &\|\bg(\theta)\|_2^2&=MN,
\end{align}
and $|\psi(\vartheta,\varphi)|\le MN$.

\begin{lemma}[Finite separation binary bound]
\label{lem:nonlocal}
For $L$ independent snapshots, let
\begin{align}
    t(\vartheta,\varphi)
    &\triangleq
    \frac{(\eta MN)^2}{1+\eta MN}
    \left(1-\frac{|\psi(\vartheta,\varphi)|^2}{(MN)^2}\right),
    \nonumber\\
    r(\vartheta,\varphi)
    &\triangleq
    1+\frac{t}{2}+\sqrt{t+\frac{t^2}{4}}.
    \label{eq:binary_eigenvalue}
\end{align}
Then
\begin{align}
    P_e^{\mathrm{iso}}(\vartheta,\varphi)
    &=I_{\frac{1}{1+r(\vartheta,\varphi)}}(L,L),
    \nonumber\\
    P_e(\bphi,\bphi+h\be_k)
    &\ge P_e^{\mathrm{iso}}(\vartheta,\varphi),
    \label{eq:nonlocal_binary}
\end{align}
where $I_x(p,q)$ is the regularized incomplete beta function.
\end{lemma}

Since $0\le|\psi(\vartheta,\varphi)|\le MN$, we have
$1\le r(\vartheta,\varphi)\le1+\eta MN$.
Monotonicity of $I_x(L,L)$ in $x$ therefore gives
\begin{align}
    P_e(\bphi,\bphi+h\be_k)
    &\ge P_\perp,
    &P_\perp&\triangleq I_{\frac{1}{2+\eta MN}}(L,L)
    \label{eq:uniform_binary}
\end{align}
for every admissible coordinate displacement.
The isolated target error probability equals $P_\perp$
when $\psi(\vartheta,\varphi)=0$; hence $P_\perp$ is the
tightest uniform lower bound obtained from the isolated
target test whenever the angular support contains such a pair.

For $\bdelta=h\be_k$, the overlapping ordered supports
have prior mass
\begin{align}
    W(h\be_k)
    &=\int_{\cS_{h\be_k}}f_{\btheta}(\bphi)\,d\bphi
    =\left(1-\frac{h}{\zeta}\right)_+^K.
    \label{eq:coordinate_overlap}
\end{align}
Using~\eqref{eq:uniform_binary} and
\eqref{eq:coordinate_overlap} in~\eqref{eq:coordinate_zzb}
and integrating over $0\le h\le\zeta$ yields,
for every coordinate,
\begin{align}
    \be_k^T\bR_e\be_k
    &\ge B_{\mathrm{FS}},
    &B_{\mathrm{FS}}
    &\triangleq
    \frac{\zeta^2P_\perp}{(K+1)(K+2)}.
    \label{eq:BNL}
\end{align}

\subsection{Local Bound from Fisher Information}
\label{subsec:local}

For one snapshot, the FIM
$\bJ_1(\btheta)\in\bbR^{K\times K}$
associated with~\eqref{eq:marginal_covariance}
is~\cite{Stoica2001}
\begin{align}
    [\bJ_1(\btheta)]_{ij}
    &=\tr{\bC^{-1}\bC_i\bC^{-1}\bC_j},
    \label{eq:fim_direct}
\end{align}
where $\bC_i\triangleq\partial\bC/\partial\theta_i$ satisfies
\begin{align}
    \bC_i
    &=\eta\bigl(\dot{\bg}_i\bg_i^H+\bg_i\dot{\bg}_i^H\bigr),
\end{align}
with $\bg_i\triangleq\bg(\theta_i)$ and
$\dot{\bg}_i\triangleq\left.\partial\bg(\theta)/\partial\theta\right|_{\theta=\theta_i}$.
Differentiating~\eqref{eq:mimo_signature} gives
\begin{align}
    \dot{\bg}(\theta)
    &=\dot{\bv}(\theta)\otimes\ba(\theta)
    +\bv(\theta)\otimes\dot{\ba}(\theta),
    \label{eq:g_derivative}
\end{align}
where $\dot{\ba}(\theta)=j\pi\cos\theta\,
\diag(0,\ldots,M-1)\ba(\theta)$, with the analogous expression
for $\dot{\bv}(\theta)$. For $L$ independent snapshots,
$\bJ_L(\btheta)=L\bJ_1(\btheta)$. Define
\begin{align}
    \overline{\bJ}_L
    &\triangleq\E_{\btheta}[\bJ_L(\btheta)].
    \label{eq:average_fim}
\end{align}
For the stated ULAs with $\eta>0$ and $MN>1$,
the continuous ordered prior ensures
$\overline{\bJ}_L\succ\mathbf0$, as required
for the inverse in the local correction.
At regular parameter values, the pointwise stochastic CRB for
locally unbiased estimators is $\bJ_L^{-1}(\btheta)$~\cite{Stoica2001}.
When its prior average is finite, define
\begin{align}
    B_{\mathrm{CRB}}^{\mathrm{avg}}
    &\triangleq\frac1K\E_{\btheta}\!\left[\tr{\bJ_L^{-1}(\btheta)}\right].
    \label{eq:average_conditional_crb}
\end{align}
By convexity of matrix inversion,
\begin{align}
    B_{\mathrm{FIM}}^{\mathrm{ref}}
    &\triangleq\frac1K\tr{\overline{\bJ}_L^{-1}}
    \le B_{\mathrm{CRB}}^{\mathrm{avg}},
    \label{eq:local_crb}
\end{align}
whenever $B_{\mathrm{CRB}}^{\mathrm{avg}}$ is finite.
The averaged FIM reference $B_{\mathrm{FIM}}^{\mathrm{ref}}$
is neither the average pointwise CRB nor a standalone Bayesian bound.

The ordered prior allows arbitrarily small target spacings. Let
$\mathbf A(\theta)\triangleq\bg(\theta)\bg^H(\theta)$ and
$D_i\triangleq\theta_{i+1}-\theta_i$.
\begin{lemma}[Small spacing Fisher degeneracy]
\label{lem:collision}
For $K\ge2$, $MN>1$, and $\eta>0$, define
$A_2\triangleq\sup_{\theta\in[\theta_{\min},\theta_{\max}]}
\|\mathbf A''(\theta)\|_F<\infty$ and
$\mathbf q_i\triangleq(\be_i-\be_{i+1})/\sqrt2$.
For $i=1,\ldots,K-1$,
\begin{align}
    D_i=0&\ \Rightarrow\ \bJ_L\mathbf q_i=\mathbf0,
    \qquad
    \mathbf q_i^T\bJ_L\mathbf q_i
    \le\frac{L\eta^2A_2^2}{2}D_i^2,
    \label{eq:collision_info_bound}\\[-1mm]
    \tr{\bJ_L^{-1}}
    &\ge\frac{2}{L\eta^2A_2^2D_i^2}
    \quad(\bJ_L\succ0).
    \label{eq:collision_crb_bound}
\end{align}
If $\bJ_L(\btheta)\succ\mathbf0$ almost everywhere
on $\cS$, then
$\E_{\btheta}[\tr{\bJ_L^{-1}(\btheta)}]=\infty$
under the ordered uniform prior without a minimum spacing.
For the stated ULAs, $K\le M+N-2$ is sufficient
for this condition.
\end{lemma}
Thus the Fisher information in the differential direction
$\mathbf q_i$ is $O(D_i^2)$ near a collision, and the pointwise
CRB grows at least as $D_i^{-2}$. This nonintegrability reflects
the loss of identifiability at coincident directions; it does not
imply divergent Bayesian MSE. Imposing a positive minimum spacing
makes $B_{\mathrm{CRB}}^{\mathrm{avg}}$ finite if $\bJ_L$
is nonsingular throughout the resulting compact support.

For $\bdelta\in\bbR^K$, define
\begin{align}
    \mathfrak r(\bdelta)
    &\triangleq
    (-\delta_1)_+
    +\sum_{i=1}^{K-1}(\delta_i-\delta_{i+1})_+
    +(\delta_K)_+,
    \label{eq:rdelta}\\
    W(\bdelta)
    &\triangleq
    \int_{\cS_{\bdelta}}f_{\btheta}(\bphi)\,d\bphi
    =\left(1-\frac{\mathfrak r(\bdelta)}{\zeta}\right)_+^K.
    \label{eq:general_overlap}
\end{align}
Thus $W(\bdelta)$ is the overlap probability of the
ordered prior supports. Set
\begin{align}
    \beta_{\rm G}(x)
    &\triangleq\frac12\left[1-\sqrt{1-e^{-x/4}}\right],
    \qquad x\ge0.
    \label{eq:beta}
\end{align}
\begin{lemma}[Local binary bound]
\label{lem:local}
For every $\bdelta\in\bbR^K$,
\begin{align}
    \int_{\cS_{\bdelta}}
    f_{\btheta}(\bphi)P_e(\bphi,\bphi+\bdelta)\,d\bphi
    &\ge W(\bdelta)\,
    \beta_{\rm G}\!\left(
    \frac{\bdelta^T\overline{\bJ}_L\bdelta}{W(\bdelta)}
    \right),
    \label{eq:local_lemma}
\end{align}
where the right side is zero when $W(\bdelta)=0$.
\end{lemma}
This inequality accounts for both the Fisher information
along the displacement and the reduction in prior overlap.

For coordinate $k$, let
$\bar v_{k,L}\triangleq[\overline{\bJ}_L^{-1}]_{kk}$ and choose
\begin{align}
    \mathbf d_k
    &\triangleq
    \frac{\overline{\bJ}_L^{-1}\be_k}{\bar v_{k,L}}.
    \label{eq:local_direction}
\end{align}
Then $\be_k^T\mathbf d_k=1$, and $h\mathbf d_k$ minimizes
$\bdelta^T\overline{\bJ}_L\bdelta$ subject to
$\delta_k=h$, attaining $h^2/\bar v_{k,L}$.
For $h\ge0$, define
\begin{align}
    c_k&\triangleq\mathfrak r(\mathbf d_k),&
    W_{S,k}(h)&\triangleq
    \left(1-\frac{c_kh}{\zeta}\right)_+^K.
    \label{eq:local_support}
\end{align}
By Lemma~\ref{lem:local}, the corresponding lower bound
on the integrated binary error is
\begin{align}
    L_k(h)
    &\triangleq W_{S,k}(h)\,
    \beta_{\rm G}\!\left(
    \frac{h^2}{\bar v_{k,L}W_{S,k}(h)}
    \right),
    \label{eq:local_candidate}
\end{align}
with $L_k(h)=0$ when $W_{S,k}(h)=0$.

\subsection{Bound Derivation and Asymptotic Behavior}
\label{subsec:combined}

The displacements $h\be_k$ and $h\mathbf d_k$ used in
Subsections~\ref{subsec:nonlocal} and~\ref{subsec:local} are both
feasible in~\eqref{eq:coordinate_zzb}. Define the finite separation
integrand bound
\begin{align}
    A_k(h)
    &\triangleq P_\perp\left(1-\frac{h}{\zeta}\right)_+^K,
    &0\le A_k(h)\le P_\perp.
    \label{eq:A_candidate}
\end{align}
The maximized integrand in~\eqref{eq:coordinate_zzb} is at least
$\max\{A_k(h),L_k(h)\}$. Since $A_k(h)\le P_\perp$,
\begin{align}
    \max\{A_k(h),L_k(h)\}
    &\ge A_k(h)+[L_k(h)-P_\perp]_+.
    \label{eq:integrand_combination}
\end{align}
Thus a local contribution is added only where it exceeds the
uniform binary error bound.

To obtain a closed-form correction, define
\begin{align}
    h_{c,k}
    &\triangleq 2\sqrt{\bar v_{k,L}}
    \sqrt{-\ln\!\left[4P_\perp(1-P_\perp)\right]},
    \nonumber\\
    \alpha_k
    &\triangleq\left(1-\frac{c_kh_{c,k}}{\zeta}\right)_+^K.
    \label{eq:hc_alpha}
\end{align}
For $\eta>0$, $0<P_\perp<1/2$. On $0\le h\le h_{c,k}$,
$W_{S,k}(h)\ge\alpha_k$, and monotonicity of
$w\beta_{\rm G}(x/w)$ in $w>0$ gives
\begin{align}
    L_k(h)
    &\ge\alpha_k\,
    \beta_{\rm G}\!\left(\frac{h^2}{\alpha_k\bar v_{k,L}}\right),
    \label{eq:local_surrogate}
\end{align}
with the right side set to zero when $\alpha_k=0$.
This surrogate exceeds $P_\perp$ on a nonempty interval
if and only if $\alpha_k>2P_\perp$. In this case, let
\begin{align}
    q_k&\triangleq\frac{P_\perp}{\alpha_k},
    &y_k&\triangleq1-2q_k,
    \label{eq:q_y}
\end{align}
and define
\begin{align}
    \Omega(y)
    &\triangleq 2y+(1-y)\ln(1-y)
    -(1+y)\ln(1+y),
    \label{eq:Omega}
\end{align}
where $\Omega(1)=2(1-\ln2)$ by continuity.
Integrating the positive part of the surrogate in
\eqref{eq:local_surrogate} minus $P_\perp$ over
$0\le h\le h_{c,k}$ yields
\begin{align}
    I_k^{\rm G}
    &\triangleq
    \begin{cases}
        \alpha_k^2\bar v_{k,L}\,\Omega(y_k),
        &\alpha_k>2P_\perp,\\
        0,&\alpha_k\le2P_\perp.
    \end{cases}
    \label{eq:local_correction}
\end{align}

Combining~\eqref{eq:BNL} and~\eqref{eq:local_correction}
establishes Proposition~\ref{prop:two_term}. Once
$\overline{\bJ}_L$ is evaluated, the bound requires
no numerical displacement optimization.

As $\eta\to0$, $P_\perp\to1/2$ and the local correction vanishes.
We define the a priori bound (APB) as the zero SNR limit
of the coordinate ZZB:
\begin{align}
    B_{\mathrm{APB}}
    &\triangleq\lim_{\eta\to0}B_{\mathrm{ZZB}}
    =\frac{\zeta^2}{2(K+1)(K+2)}.
    \label{eq:apb}
\end{align}
This differs from the average prior variance for $K>1$.
To characterize the gap, set $\theta_0=\theta_{\min}$ and
$\theta_{K+1}=\theta_{\max}$, and let $\btheta_{-k}$ contain
the other directions. By~\eqref{eq:ordered_prior},
$\theta_k\mid\btheta_{-k}\sim
\mathcal U(\theta_{k-1},\theta_{k+1})$.
Moreover, $\mathfrak r(\bdelta)\ge h$ for $\delta_k=h$,
so $h\be_k$ maximizes the overlap at zero SNR.
The law of total variance therefore gives
\begin{align}
    B_{\mathrm{APB}}
    &=\frac1K\sum_{k=1}^K
    \E[\operatorname{Var}(\theta_k\mid\btheta_{-k})],
    \nonumber\\
    \frac{\zeta^2}{6(K+1)}-B_{\mathrm{APB}}
    &=\frac1K\sum_{k=1}^K
    \operatorname{Var}(\E[\theta_k\mid\btheta_{-k}])
    \nonumber\\
    &=\frac{\zeta^2(K-1)}{6(K+1)(K+2)}.
    \label{eq:apb_gap}
\end{align}
This gap is due to variation in the conditional means and is
intrinsic to the coordinate ZZB, not the two-term relaxation.
It vanishes for $K=1$.

For fixed $L$, $M$, and $N$, as $\eta\to\infty$,
\begin{align}
    P_\perp
    &\sim\binom{2L-1}{L-1}(\eta MN)^{-L}.
    \label{eq:nonlocal_diversity}
\end{align}
Thus the finite separation term decays with snapshot diversity
order $L$. Furthermore,
$-\ln[4P_\perp(1-P_\perp)]=L\ln\eta+O(1)$.
If $c_k^2\bar v_{k,L}\ln\eta\to0$, then
$h_{c,k}\to0$ and $\alpha_k,y_k\to1$, giving
\begin{align}
    I_k^{\rm G}
    &\sim2(1-\ln2)\bar v_{k,L}.
    \label{eq:gaussian_local_asymptotic}
\end{align}
Hence the local correction scales with the corresponding diagonal
entry of $\overline{\bJ}_L^{-1}$, with asymptotic factor
$2(1-\ln2)\simeq0.6137$.

\section{Numerical Results}
\label{sec:numerical}

Unless stated otherwise, $M=N=15$, $L=10$, and
$[\theta_{\min},\theta_{\max}]=[-60^\circ,60^\circ]$.
We use $\mathrm{SNR}=10\log_{10}\eta$ and report root mean
square error (RMSE) in degrees, with all computations performed
in radians. The prior averaged FIM $\overline{\bJ}_L$ is
computed numerically; the two terms of~\eqref{eq:final_bound}
are then evaluated in closed form.

For the experiments with a minimum target spacing
$\theta_{k+1}-\theta_k\ge\Delta_{\min}$, where
$(K-1)\Delta_{\min}<\zeta$, the variables
$\widetilde\theta_k=\theta_k-(k-1)\Delta_{\min}$
follow an ordered uniform prior of width
$\zeta_{\rm eff}=\zeta-(K-1)\Delta_{\min}$.
We evaluate the steering vectors and FIM at the physical
angles $\theta_k$, recompute the prior averaged FIM, and use
$\zeta_{\rm eff}$ in the overlap probabilities and the
bound. The separated prior uses $\Delta_{\min}=10^\circ$.

\subsection{Validation Against the Vector ZZB and Snapshot Dependence}

Fig.~\ref{fig:L_decomposition} compares the proposed bound
with a numerical evaluation of~\eqref{eq:coordinate_zzb}
for $K=2$ and $L\in\{5,10,40\}$. 
Unlike the proposed analytical bound, the numerical
reference applies valley filling to the optimized
binary error profile before integration. 
The two-term bound closely follows this
reference at low SNR but is more conservative in the
high SNR local regime. The finite separation term dominates
at low SNR, whereas the local correction determines the
high SNR behavior. Increasing $L$ shifts the transition to
lower SNR and reduces both contributions, consistent with
\eqref{eq:nonlocal_diversity} and
$\overline{\bJ}_L=L\overline{\bJ}_1$.

Because direct vector ZZB evaluation requires multidimensional
prior integration and displacement optimization, this comparison
is restricted to $K=2$. The remaining experiments use the
two-term bound and show $B_{\mathrm{FIM}}^{\mathrm{ref}}$
only as a local information reference.

\begin{figure}[t]
    \centering
    \input{Figures/L_decomposition_new_bound}
    \caption{Numerical vector ZZB comparison and bound decomposition
    for $K=2$ and $L\in\{5,10,40\}$. Dotted with circles:
    vector ZZB with valley filling; solid: two-term bound;
    dashed: finite separation term; dash-dotted: local correction.
    Colors indicate $L$.}
    \label{fig:L_decomposition}
\end{figure}

\subsection{Effect of Target Spacing}

Fig.~\ref{fig:estimator_results} considers $K=5$ with
$\Delta_{\min}=0^\circ$ and $10^\circ$. Under the
unrestricted ordered uniform prior, the probability that the
smallest adjacent spacing is below $d$, for
$(K-1)d<\zeta$, is
\begin{align}
    \Pr\!\left\{\min_i(\theta_{i+1}-\theta_i)<d\right\}
    &=1-\left(1-\frac{(K-1)d}{\zeta}\right)^K.
    \label{eq:min_gap_probability}
\end{align}
For $\zeta=120^\circ$ and $K=5$, this probability is
$86.8\%$ at $d=10^\circ$, illustrating the prevalence
of closely spaced targets. By Lemma~\ref{lem:collision},
the prior averaged pointwise CRB diverges without a
spacing constraint, although the averaged FIM reference
remains finite.

The figure also reports the RMSE of a stochastic maximum
likelihood (ML) estimator initialized by multiple signal
classification (MUSIC)~\cite{Schmidt1986}, using
$10^4$ trials per SNR.
The numerical likelihood maximization is not guaranteed to
reach the global optimum. Without a spacing constraint, its RMSE remains much
larger than the local information reference at moderate
and high SNR. With
$\Delta_{\min}=10^\circ$, the estimator approaches this
reference at high SNR. In this case, the prior averaged
pointwise CRB is finite provided the FIM remains
nonsingular throughout the constrained support.
The spacing constraint also reduces the effective prior
width from $120^\circ$ to $80^\circ$ and hence lowers the APB.

\begin{figure}[t]
    \centering
    \input{Figures/estimator_results_new_bound}
    \caption{RMSE for $K=5$ and $\Delta_{\min}=0^\circ$
    (blue) or $10^\circ$ (orange). Solid: two-term ZZB;
    dashed: averaged FIM reference; dotted: APB;
    dash-dotted: stochastic ML initialized by MUSIC
    ($10^4$ trials per SNR).}
    \label{fig:estimator_results}
\end{figure}

\subsection{Transmit and Receive Allocation at Fixed \texorpdfstring{$MN$}{MN}}

For $K=2$ and $MN=225$, Fig.~\ref{fig:fixed_mn}
compares five antenna allocations under equal total
transmit energy. Since $N$ varies, we impose
$\int_{\mathcal T}\bs(t)\bs^H(t)\,dt=\bI_N/N$,
corresponding to $\eta_{\rm eff}=\eta/N$ in
\eqref{eq:marginal_covariance}, to ensure a common
transmit energy budget. At low SNR, the balanced
allocation $(15,15)$ yields the smallest bound,
as receive array gain improves discrimination
between separated target configurations. At high
SNR, the finite separation contribution diminishes,
and the substantially larger virtual aperture of
$(1,225)$ outweighs its reduced receive gain,
yielding the smallest bound. Intermediate allocations
exhibit the transition between these regimes.
This illustrates how the two contributions capture
the competing effects of receive gain and angular
resolution across SNR.

\begin{figure}[t]
    \centering
    \input{Figures/fixed_MN_new_bound}
    \caption{RMSE for $K=2$, $MN=225$, and equal total
    transmit energy. Colors identify $(M,N)$; solid:
    two-term ZZB; dashed: averaged FIM reference;
    black dotted: APB.}
    \label{fig:fixed_mn}
\end{figure}

\section{Conclusion}

The proposed two-term ZZB provides a rigorous Bayesian MSE benchmark for multitarget DoA estimation under a prior permitting arbitrarily close directions. It combines a finite separation contribution with a local correction retaining target coupling and virtual array geometry. The collision analysis explains why the prior averaged pointwise CRB diverges although Bayesian MSE remains finite. Once the prior averaged FIM is computed, the bound enables repeated comparisons of snapshot counts, spacing assumptions, and transmit and receive allocations without displacement optimization or binary error integration. Numerical results show close agreement with the vector ZZB at low SNR and a more conservative high SNR bound. This explicit benchmark thus trades some tightness for computational accessibility without sacrificing lower-bound validity.

\appendix[Proofs]

\emph{Proof of Lemma~\ref{lem:nonlocal}.}
Let $a\triangleq\eta MN$ and $c\triangleq|\psi(\vartheta,\varphi)|/(MN)$.
The two covariance matrices in~\eqref{eq:isolated_pair} have equal
determinants $(1+a)$, and
\begin{align}
\tr{\bSigma_0^{-1}\bSigma_1}
=MN+\frac{a^2(1-c^2)}{1+a}.
\label{eq:pair_trace_app}
\end{align}
Since $\bSigma_0^{-1/2}\bSigma_1\bSigma_0^{-1/2}$ differs from the
identity on a subspace of dimension at most two, its potentially nonunit eigenvalues
are $r$ and $r^{-1}$. Their product is one and their sum is $2+t$,
which gives~\eqref{eq:binary_eigenvalue}. After whitening under
$\mathcal H_0$ and diagonalizing the generalized covariance pair,
the log-likelihood ratio selects $\mathcal H_1$ when $U>rV$, where
$U,V$ are independent $\operatorname{Gamma}(L,1)$ energies under
$\mathcal H_0$. The two conditional error probabilities coincide, 
and $U/(U+V)\sim\operatorname{Beta}(L,L)$ gives
$P_e^{\mathrm{iso}}=\Pr\{U>rV\}=I_{1/(1+r)}(L,L)$.
The full observation adds a contribution from unchanged targets,
independent of the isolated target and identical
in distribution under both hypotheses. Data processing gives~\eqref{eq:nonlocal_binary}.
\hfill$\square$

\noindent\emph{Proof of Lemma~\ref{lem:collision}.}
Let $\mathbf A(\theta)=\bg(\theta)\bg^H(\theta)$, so that $\bC_i=\eta\mathbf A'(\theta_i)$. For $\mathbf q_i=(\be_i-\be_{i+1})/\sqrt2$,~\eqref{eq:fim_direct} gives
\begin{align}
\mathbf q_i^T\bJ_L\mathbf q_i
&=\frac{L}{2}\left\|\bC^{-1/2}(\bC_i-\bC_{i+1})\bC^{-1/2}\right\|_F^2.
\label{eq:collision_energy_app}
\end{align}
At $D_i=0$, this vanishes and, since $\bJ_L\succeq\mathbf0$, $\bJ_L\mathbf q_i=\mathbf0$. For $D_i>0$, $\bC\succeq\bI_{MN}$ and the mean value theorem give $\|\bC_i-\bC_{i+1}\|_F\le\eta A_2D_i$, proving~\eqref{eq:collision_info_bound}. If $\bJ_L\succ\mathbf0$, Cauchy--Schwarz gives $(\mathbf q_i^T\bJ_L\mathbf q_i)(\mathbf q_i^T\bJ_L^{-1}\mathbf q_i)\ge1$; with $\tr{\bJ_L^{-1}}\ge\mathbf q_i^T\bJ_L^{-1}\mathbf q_i$, this yields~\eqref{eq:collision_crb_bound}. Under~\eqref{eq:ordered_prior}, each adjacent spacing has density $f_{D_i}(d)=K\zeta^{-1}(1-d/\zeta)^{K-1}$ for $0<d<\zeta$. Since $f_{D_i}(0)>0$,~\eqref{eq:collision_crb_bound} gives a divergent integral proportional to $\int_0^\epsilon d^{-2}f_{D_i}(d)\,dd$.

For the local behavior, set $\theta_i=\mu-d/2$ and $\theta_{i+1}=\mu+d/2$, with $\bC_0$ denoting the covariance at $d=0$. Smoothness gives $\bC_i-\bC_{i+1}=-\eta d\mathbf A''(\mu)+O(d^3)$ and $\bC=\bC_0+O(d^2)$, hence
\begin{align}
\mathbf q_i^T\bJ_L\mathbf q_i
&=\frac{L\eta^2d^2}{2}
\left\|\bC_0^{-1/2}\mathbf A''(\mu)\bC_0^{-1/2}\right\|_F^2+O(d^4).
\end{align}
Thus the separation information has quadratic leading behavior whenever the displayed coefficient is nonzero. \hfill$\square$

\noindent\emph{Proof of Lemma~\ref{lem:local}.}
The $K+1$ gaps of an ordered vector, including both boundary
gaps, are nonnegative and sum to $\zeta$.
Requiring $\bphi+\bdelta\in\cS$ imposes lower limits
$(-\delta_1)_+$, $(\delta_i-\delta_{i+1})_+$, and
$(\delta_K)_+$ on these gaps. Their sum is
$\mathfrak r(\bdelta)$, leaving a simplex of width
$(\zeta-\mathfrak r(\bdelta))_+$.
The volume ratio gives~\eqref{eq:general_overlap};
$\bdelta=h\be_k$ gives~\eqref{eq:coordinate_overlap}.

Let $\rho(\bphi,\bdelta)$ be the Bhattacharyya coefficient of the two $L$-snapshot hypotheses. For equal priors~\cite{Sason2015},
\begin{align}
P_e(\bphi,\bphi+\bdelta)
\ge\frac12\left[1-\sqrt{1-\rho^2(\bphi,\bdelta)}\right].
\label{eq:tv_bc_app}
\end{align}
With $\bC_0=\bC(\bphi)$, $\bC_1=\bC(\bphi+\bdelta)$, and $e^{t_i}$ the eigenvalues of $\bC_0^{-1/2}\bC_1\bC_0^{-1/2}$, direct Gaussian integration and $\ln\cosh u\le u^2/2$ give
\begin{align}
\rho
=\prod_i\operatorname{sech}^{L}\!\left(\frac{t_i}{2}\right)
\ge e^{-d_{\rm G}^2/8},
\label{eq:gaussian_bc_app}
\end{align}
where $d_{\rm G}^2=L\|\log(\bC_0^{-1/2}\bC_1\bC_0^{-1/2})\|_F^2$. The affine invariant covariance distance is bounded by the
metric length of the covariance path~\cite{Amari2016};
Cauchy--Schwarz bounds its square by the Fisher energy:
\begin{align}
 d_{\rm G}^2
 &\le D_L(\bphi,\bdelta)
 \triangleq\int_0^1\bdelta^T\bJ_L(\bphi+t\bdelta)\bdelta\,dt.
 \label{eq:path_energy_app}
\end{align}
Thus $P_e\ge\beta_{\rm G}(D_L)$. Convexity ensures
$\cS_{\bdelta}+t\bdelta\subseteq\cS$ for $t\in[0,1]$.
Using the constant prior density and $\bJ_L\succeq\mathbf0$ gives
\begin{align}
\int_{\cS_{\bdelta}}f_{\btheta}(\bphi)D_L(\bphi,\bdelta)\,d\bphi
\le\bdelta^T\overline{\bJ}_L\bdelta.
\end{align}
Since $\beta_{\rm G}$ is decreasing and convex and the prior mass of $\cS_{\bdelta}$ is $W(\bdelta)$, conditional Jensen followed by monotonicity gives~\eqref{eq:local_lemma}; for $W(\bdelta)=0$, both sides are zero. \hfill$\square$

\noindent\emph{Proof of Proposition~\ref{prop:two_term}.}
Since $A_k$ and $L_k$ arise from feasible displacements in~\eqref{eq:coordinate_zzb} and $A_k\le P_\perp$,
$\max\{A_k,L_k\}\ge A_k+[L_k-P_\perp]_+$. For $h\le h_{c,k}$, $W_{S,k}(h)\ge\alpha_k$; moreover $w\beta_{\rm G}(x/w)$ is nondecreasing since, with $z=x/w$, its derivative is $\beta_{\rm G}(z)-z\beta_{\rm G}'(z)\ge0$. For $\alpha_k=0$, the correction vanishes. Otherwise,
$L_k(h)\ge\alpha_k\beta_{\rm G}(h^2/(\alpha_k\bar v_{k,L}))$.
If $\alpha_k>2P_\perp$, this surrogate exceeds $P_\perp$
for $0\le h<h_{*,k}$, where
$h_{*,k}=2\sqrt{\alpha_k\bar v_{k,L}}\sqrt{-\ln(1-y_k^2)}\le h_{c,k}$.
The inequality follows from $\alpha_k\le1$ and
$y_k\le1-2P_\perp$. With
$u=\sqrt{1-e^{-h^2/(4\alpha_k\bar v_{k,L})}}$,
\begin{align*}
\int_0^{h_{*,k}}h\!\left[\alpha_k\beta_{\rm G}\!\left(\frac{h^2}{\alpha_k\bar v_{k,L}}\right)-P_\perp\right]dh
=\alpha_k^2\bar v_{k,L}\Omega(y_k),
\end{align*}
with zero correction otherwise. Integrating $A_k(h)$ over $0\le h\le\zeta$ gives~\eqref{eq:BNL}; combining both contributions and averaging over $k$ proves~\eqref{eq:final_bound}. \hfill$\square$

\bibliographystyle{IEEEtran}
\bibliography{bibliography}

\end{document}